\documentclass[12pt]{article}
\begin{document}

\begin{titlepage}
\centerline{\bf Quasi-multi-Regge processes
with  a quark exchange in the $t$-channel}
\date{}
\begin{center}
L. N. Lipatov {$^{\dagger}$\\
Universite de Montpellier-2\\
34095, Montpellier, cedex 05, France\\
and\\
Petersburg Nuclear Physics Institute \\
Gatchina, 188300, St. Petersburg, Russia}
\end{center}
\begin{center}
M. I. Vyazovsky \\
St. Petersburg State University\\
St. Petersburg, Russia
\end{center}
\vskip 15.0pt
\centerline{\bf Abstract}
\noindent
The QCD amplitudes for  particle's production in
the quasi-multi-Regge kinematics with  a quark exchange in
crossing channels are calculated in the Born
approximation. In particular they are needed  to find
next-to-leading corrections to the quark Regge trajectory and
to the integral kernel of the Bethe-Salpeter equation for
the $t$-channel partial wave with  fermion quantum numbers and
a negative signature. The
gauge-invariant  action for the interaction of the
reggeized quarks and gluons with the usual particles is
constructed.

\vskip 3cm \hrule \vskip 3cm

\noindent

\noindent ${\dagger}$ {\it Supported by CRDF
  RP1-2108 and INTAS  97-31696 grants}

\vfill

\end{titlepage}

\section{Introduction}
It is well known \cite{bfkl} \cite{s2}, that  gluon is reggeized
in the
framework
of the perturbative QCD. Namely, the scattering amplitude for the
process $A+B \rightarrow A'+B'$ at large
energies $s^{1/2}$ ($s=(p_A+p_B)^2$) and fixed momentum transfers
$q=(-t)^{1/2}$ ($t=(p_A-p_B)^2$) has the
Regge form
\begin{equation}
f(s, t)=f_0 (s,t)\,\, \Gamma _A^{A'}
(q^2)\,\left(\frac{s}{q^2}\right)^{\omega
(q^2)} \Gamma _B^{B'} (q^2)\,,
\end{equation}
where
\begin{equation}
f_0 (s,t) = T_A^{A'}(c) \,\frac{2\,s}{t}\,T_B^{B'}(c)\,\,,
\,\,\,[T(a),\,T(b)]=i\,f_{abc}T(c)\,.
\end{equation}
The quantities $\Gamma_A^{A'}(q^2)$ and $j=1+\omega (q^2)$ are
the
reggeon-particle vertex and the gluon Regge trajectory
correspondingly.
They are known up to the second order of the
perturbation
theory \cite{b2}. The BFKL Pomeron in QCD is a compound state of two
reggeized gluons in the leading and next-to-leading approximations
\cite{bfkl}, \cite{b3}. With the use of  non-abelian
renormalization
schemes and the BLM procedure for the  scale setting of the
running QCD coupling constant
the
prediction of the BFKL theory in the next-to-leading
approximation
turns out to be in an agreement with the experimental data of the L3 group
on the hadron production in the virtual photon-photon collisions
\cite{b4}.

There are  other colourless objects  which can be constructed
as composite states of several reggeized gluons in the leading
logarithmic approximation (LLA). The simplest state of
such
type is the Odderon having the charge parity $C=-1$ and the
signature
$P_j=-1$. These quantum numbers are opposite in sign to that for the
Pomeron. The Odderon can be constructed as a composite state of
three
reggeized gluons \cite{odd}. In LLA one can derive an
effective field theory for the reggeized gluon interactions in the
2+1 space-time \cite{s1}. The effective lagrangian for the
interaction of
the reggeized gluons with the quarks and gluons is obtained from
QCD \cite{eff}.

The quark in QCD is also reggeized in the leading
logarithmic approximation (LLA) \cite{fsh} (for the case of the
electron in QED it was discovered in ref. \cite{ggmz}). Indeed, one
can show, that in the $t$-channel partial wave with the positive
signature there is a Regge pole with
its trajectory going through the
point $j=1/2$ for $t=M^2$. Further, for the amplitude with a negative
signature in ref \cite{fsh} the integral equation similar to the
BFKL equation was obtained. Because all known mesons and barions are
compound states of quarks, one should construct the theory of the
reggeized quark interactions. In particular, it is needed to
calculate  next-to-leading corrections to the quark Regge
trajectory using the methods which were developed earlier for the
gluodynamics \cite{bfkl}, \cite{b89}. These methods are based on an
intensive use of dispersion relations and unitarity
conditions starting from the Born expression for the inelasic
amplitudes in the quasi-multi-Regge kinematics. An alternative method,
based on the effective action, can be also used \cite{eff}.

In refs. \cite{s1}, \cite{fsh}  simple rules for finding the Born
amplitudes
in the multi-Regge kinematics were formulated. To calculate the loop
corrections to LLA one should know  inelastic amplitudes in the
quasi-multi-Regge kinematics for the case, when all pair energies except
one are
large in comparison with momentum transfers. Such amplitudes for
the processes with the gluon exchange in the $t$-channel
are known (see \cite{b89}).

This paper is devoted to the calculation of the Born amplitudes
for the quasi-multi-Regge kinematics with
the quark exchange in the crossing channel. It is well known, that
the Regge
trajectory is universal. Thus, it is enough to consider only one
process with the fermion exchange: for example, the collision of
quark and anti-quark. In Section 2 we calculate all amplitudes
neccessary for finding two-loop corrections to the Regge trajectory
of the quark and one-loop corrections to the integral kernel for the
amplitude with the negative signature.

In Section 3 the effective action describing the interaction of the
reggeized quark with usual quarks and gluons is suggested. From this
action all results of Section 2 can be derived. In the end of
Section 3 the program for the calculation of  next-to-leading
corrections to the quark Regge trajectory  and to various reggeon
vertices is formulated.

\section{Inelastic processes in the quasi-multi-Regge
kinematics}
We consider the Yang-Mills theory with the gauge group $SU(N_c)$
containing also  the quark field in the fundamental
representation.
The generators $T^{a}$ satisfy
the commutation relation $[T^{a},T^{b}]=if^{abc}T^{c}$ and the
normalization condition $tr(T^a T^b)=\frac{1}{2}\delta^{ab}$.

We introduce two light-cone vectors
corresponding to the massless particle momenta in the c.m. system:
$p_{1}=(p,0,0,p)$ and $p_{2}=(p,0,0,-p)$ and define
light-cone components of the Lorentz vector $a_{\mu}$ as follows
$a^{\pm}=a_{\mp}=a_{\mu} n^{\pm}_{\mu}=a^{0}\pm a^{3}$,
where $n^{+}=p_{2}/p,\,
n^{-}=p_{1}/p$.
For  the scalar products of two vectors
the sum over the Lorentz indices is implied:
$ab=a_{\mu}b_{\mu}\equiv a_{\mu}b_{\nu}g^{\mu\nu}=
(a_{+}b_{-}+a_{-}b_{+})/2+a_{\perp}b_{\perp}\,$.

Let us consider the annihilation of the quark and
anti-quark with masses $M$
into three gluons
$A+B\to A'+A''+B'$. One can use the Sudakov decomposition for the
momenta of the colliding particles
\begin{equation}
p_{A}=p_{1}+\frac{M^{2}}{4p^{2}}\,p_{2}, \quad
p_{B}=p_{2}+\frac{M^{2}}{4p^{2}}\,p_{1},
\label{21}
\end{equation}
where $s=(p_{A}+p_{B})^{2}\simeq 4p^{2}+2M^{2}$.

In the quasi-multi-Regge  kinematics for the gluon production
in the fragmentation region of the initial quark with the momentum
$p_A$ the following relations among
invariants
\begin{eqnarray}
&& u_1=(p_{A'}-p_{B})^{2}\sim u_2=(p_{A''}-p_{B})^{2}\sim s\gg M^{2}
\nonumber \\
&& t_1=(p_{A}-p_{A'})^{2}\sim t_2=(p_{A}-p_{A''})^{2}\sim M^{2}
\label{22a} \\
&& t=(p_{B'}-p_{B})^{2}\sim M^{2},\quad
s_{1}=(p_{A'}+p_{A''})^{2}\sim M^{2} \nonumber
\end{eqnarray}
are valid.  The  components of particle momenta in this kinematics
are
\begin{eqnarray}
&& p_{B'}^+\ll p_{A'}^+\sim p_{A''}^+\, ,
\ p_{A}^+\simeq p_{A'}^++p_{A''}^+\, ,\nonumber \\
&& p_{A'}^-\sim p_{A''}^-\ll p_{B'}^-\simeq p_{B}^-\, , \label{22} \\
&& p_{A'\perp}\sim p_{A''\perp}\sim p_{B'\perp}\sim M. \nonumber
\end{eqnarray}
For the momentum transfer
$q=p_{B'}-p_{B}=p_{A}-p_{A'}-p_{A''}$ we obtain
\begin{eqnarray}
&& q^{+}\ll p_{A}^+,p_{A'}^+,p_{A''}^+\, , \nonumber \\
&& q^{-}\ll p_{B}^-,p_{B'}^-\, , \label{23} \\
&& q_{}^{2}\simeq q_{\perp}^{2}\sim M^{2}. \nonumber
\end{eqnarray}

In the Born approximation the amplitude of annihilation of the quark
$A$ and the virtual antiquark $C$ with the
momentum
$p_{C}=-q$ into  two gluons $A',A''$ equals
\begin{eqnarray}
{\cal A}(AC\to A'A'')\!\!=\!\!
g^{2}e^{*}_{A'\mu}e^{*}_{A''\nu}\cdot \bar v_{C}
\left(\gamma_{\nu}(\hat p_{A}-\hat p_{A'}-M)^{-1}\gamma_{\mu}T^{a''}T^{a'}+
\right.\nonumber \\  \left.
+\gamma_{\mu}(\hat p_{A}-\hat p_{A''}-M)^{-1}\gamma_{\nu}T^{a'}T^{a''}+
\gamma_{\mu\nu\sigma}(p_{A'},p_{A''})s_{1}^{-1}\gamma_{\sigma}
(T^{a'}T^{a''}-T^{a''}T^{a'})\right)u_{A},
\label{24}
\end{eqnarray}
where $e_{A'}$,$e_{A''}$ are polarization vectors of the final
gluons,
$a',a''$ are their colour indices and $u_{A}$, $\bar v_{C}$
are the fermion spinors. The tensor

\begin{equation}
\gamma_{\mu\nu\sigma}(p,p')=(p'-p)_{\sigma}g_{\mu\nu}-
(p+2p')_{\mu}g_{\nu\sigma}+(2p+p')_{\nu}g_{\mu\sigma}
\label{25}
\end{equation}
is the triple Yang-Mills vertex.

It turns out, that with taking into account eq. (7) the amplitude of
the process
$A+B\to A'+A''+B'$ can be
written in
in the  form
\begin{equation}
{\cal A}(AB\to A'A''B')\!=\!
ge^{*}_{A'\mu}e^{*}_{A''\nu}e^{*}_{B'\lambda}\cdot \bar v_{B}
\left(A_{\mu\nu\lambda}^{a'a''b'}+\delta_{\mu\nu\lambda}^{a'a''b'}
\right)u_{A}\,.
\label{26}
\end{equation}
Here we denoted
\begin{eqnarray}
A_{\mu\nu\lambda}^{a'a''b'}= &&
g^{2}\gamma^{(-)}_{\lambda}(-p_{B'},q)\ T^{b'}(\hat q -M)^{-1}
\left(\gamma^{(+)}_{\nu}(p_{A''},q)\ (\hat p_{A}-\hat p_{A'}-M)^{-1}
\gamma_{\mu}\ T^{a''}T^{a'}+
\right. \nonumber \\ &&
+\gamma^{(+)}_{\mu}(p_{A'},q)\ (\hat p_{A}-\hat p_{A''}-M)^{-1}
\gamma_{\nu}\ T^{a'}T^{a''}+
\nonumber \\ && \left.
+\gamma_{\mu\nu\sigma}(p_{A'},p_{A''})\, s_{1}^{-1} \,
\gamma^{(+)}_{\sigma}(p_{A'}+p_{A''},q)
\ (T^{a'}T^{a''}-T^{a''}T^{a'})\right)u_{A}
\end{eqnarray}
and introduced the effective vertices
\begin{equation}
\gamma^{(+)}_{\lambda}(p,q)=\gamma_{\lambda}+
(\hat q -M)\frac{n_{\lambda}^{+}}{p^{+}}\, ,
\ \gamma^{(-)}_{\mu}(p,q)=\gamma_{\mu}+
(\hat q -M)\frac{n_{\mu}^{-}}{p^{-}}\,.
\label{27}
\end{equation}
These vectors have the property of the transversality
with respect to the gluon momentum
$p_{\mu}\gamma^{(+)}_{\mu}(p,q)=\hat p+(\hat q -M)
\to 0$, $p_{\lambda}\gamma^{(-)}_{\lambda}(-p,q) \to 0$
providing that the colliding fermions are on their mass shell. It is
important, that they differ from the usual $\gamma$-matrices only
by the terms which are proportional to the factor $\hat q -M$
cancelling the neighbouring fermion propagator. These terms are
contributions from the crossing diagrams having the additional
propagators in the direct channel and proportional to $1/p^{\pm}$.
One can  verify, that
the term $A_{\mu\nu\lambda}$ is not gauge invariant:
\begin{eqnarray}
&& e^{*}_{A'\mu}p_{A''\nu}e^{*}_{B'\lambda}
\cdot A_{\mu\nu\lambda}^{a'a''b'}=
-g^{2}e^{*}_{B'\lambda}\gamma^{(-)}_{\lambda}(-p_{B'},q)\,T^{b'}
\,\left(\frac{e^{+*}_{A'}}{p_{A}^+}\,T^{a''}T^{a'}+
\frac{p^{+}_{A''}e^{+*}_{A'}}{p^{+}_{A'}p^{+}_{A}}\,T^{a'}T^{a''}
\right) \nonumber \\
&& e^{*}_{A'\mu}e^{*}_{A''\nu}p_{B'\lambda}
\cdot A_{\mu\nu\lambda}^{a'a''b'}= 0
\label{28}
\end{eqnarray}

The additional contribution $\delta_{\mu\nu\lambda}^{a'a''b'}$
needed for the gauge invariance of the production amplitude does
not has any pole in the $q^2$-channel and
corresponds to the Feynman diagrams having singularities in the
direct energy invariants.  It  is the sum of terms containing in the
denominator one of the following expressions
$$
d_{1}=(p_{A}p_{B})(p_{A'}p_{B})=p^2p_A^+p_{A'}^+\, , \quad
d_{2}=(p_{A}p_{B})(p_{A''}p_{B})\, , \quad
d_{3}=(p_{A'}p_{B})(p_{A''}p_{B})\,
$$
with $d_{1}^{-1}+d_{2}^{-1}=d_{3}^{-1}$. It can be written as
follows

\begin{equation}
\delta_{\mu\nu\lambda}^{a'a''b'}
=g^{2}\gamma^{(-)}_{\lambda}(-p_{B'},q)\left(
a\frac{p_{B\mu}p_{B\nu}}{d_{1}}+b\frac{p_{B\mu}p_{B\nu}}{d_{2}}
\right)\,T^{b'}T^{a''}T^{a'}
+\left(p_{A'}\leftrightarrow p_{A''},\, a'\leftrightarrow a''\right)
\label{29}
\end{equation}
with the dimensionless constants $a,b$.
To satisfy the transversality conditions one should chose the unique
values for them: $a=0,b=1$ corresponding to the correct weights for the
corresponding Feynman diagrams.

Thus, we obtain the induced vertex
\begin{equation}
\Gamma_{\mu\nu}^{a'a''}=
-g^{2}(\hat q -M)\frac{n^{+}_{\mu}n^{+}_{\nu}}{p_{A}^+}
\left(\frac{T^{a'}T^{a''}}{p_{A'}^+}+\frac{T^{a''}T^{a'}}{p_{A''}^+}
\right)
\label{210}
\end{equation}

If in the final state we have
quark $A'$, anti-quark $A''$  and gluon $B'$, the production amplitude is
constructed in a more simple way
\begin{equation}
{\cal A}(AB\to A'A''B')\!=\!
ge^{*}_{B'\lambda}\cdot \bar v_{B} T^{b'}\gamma^{(-)}_{\lambda}(-p_{B'},q)\,
(\hat q -M)^{-1}\Gamma_{q\bar q}\, u_{A},
\label{211}
\end{equation}
where
\begin{eqnarray}
\Gamma_{q\bar q} &=&
-g^{2}\left(
(\bar u_{A'}\gamma_{\sigma}T^{c}v_{A''}) s_{1}^{-1}
\gamma^{(+)}_{\sigma}(p_{A'}+p_{A''},q)T^{c}+\right.
\nonumber \\ && \left.
+(p_{A}-p_{A'})^{-2}\,T^{c}\gamma^{(+)}_{\sigma}(p_{A}-p_{A'},q) v_{A''}
\otimes \bar u_{A'}\gamma_{\sigma}T^{c}
\cdot \delta_{AA'}\delta_{BA''}\right) \nonumber
\end{eqnarray}
and the multiplier $\delta_{AA'}\delta_{BA''}$ in the last line
corresponds to the conservation of the quark flavour.

Let us consider now the more complicated process $A+B\to
A'+D_{1}+D_{2}+B'$ in the
quasi-multi-Regge kinematics where the gluons $D_1, D_2$ in the central
rapidity region are
produced with a fixed invariant mass
\begin{eqnarray}
&& p_{B'}^+\ll p_{D_{1}}^+\sim p_{D_{2}}^+\ll p_{A'}^+\simeq
p_{A}^+\, ,
\nonumber \\
&& p_{A'}^-\ll p_{D_{1}}^-\sim p_{D_{2}}^-\ll p_{B'}^-\simeq
p_{B}^-\, ,
\nonumber \\
&& p_{A'\perp}\sim p_{D_{1}\perp}\sim p_{D_{2}\perp}\sim p_{B'\perp}\sim M,
\label{212} \\
&& s_{2}=(p_{D_{1}}+p_{D_{2}})^{2}\sim M^{2}, \nonumber \\
&& t_{1}=(p_{A}-p_{A'})^{2}\sim M^{2},
t_{2}=(p_{B'}-p_{B})^{2}\sim M^{2} \,.\nonumber
\end{eqnarray}
Its amplitude has the property of the factorization
$$
{\cal A}(AB\to A'D_{1}D_{2}B')\!=\!
\Gamma(B\to B')(\hat q_{2}-M)^{-1}\cdot\Gamma_{D_{1}D_{2}}\cdot
(\hat q_{2}-M)^{-1}\Gamma(A\to A'),
$$

For the vertices corresponding to the transition of the initial
(reggeized)
fermions  $A, B$ into the final gluons $A', B'$ we have the known result
$$
\Gamma(A\to A')=
-ge^{*}_{A'\mu}\cdot
\gamma^{(+)}_{\mu}(p_{A'},q_{1})\, T^{a'}u_{A}\, ,
$$
$$
\Gamma(B\to B')=
-ge^{*}_{B'\lambda}\cdot \bar v_{B}T^{b'}
\gamma^{(-)}_{\lambda}(-p_{B'},q_{2})
$$

The effective amplitude for the gluon production in the virtual
quark-anti-quark
collisions is
\begin{eqnarray}
\Gamma_{D_{1}D_{2}}=&& -g^{2}e^{*}_{D_{1}\mu}e^{*}_{D_{2}\nu}\cdot
\left(
\gamma^{(+)}_{\nu}(p_{D_2},q_{2})(\hat q_{2}+\hat p_{D_2}-M)^{-1}
\gamma^{(-)}_{\mu}(-p_{D_1},q_{2})\, T^{d_{2}}T^{d_{1}}+
\right. \nonumber \\
&& +\gamma^{(+)}_{\nu}(p_{D_1},q_{2})(\hat q_{2}+\hat p_{D_1}-M)^{-1}
\gamma^{(-)}_{\mu}(-p_{D_2},q_{2})\, T^{d_{1}}T^{d_{2}}+ \label{213} \\
&& \left. +\gamma_{\mu\nu\sigma}(p_{D_1},p_{D_2})\, s_{2}^{-1} \,
\gamma_{\sigma}^{(+-)}(q_{1},q_{2})
\, (T^{d_{1}}T^{d_{2}}-T^{d_{2}}T^{d_{1}})
+\Delta_{\mu\nu}^{d_{1}d_{2}}(q_{1},q_{2})
\right). \nonumber
\end{eqnarray}
Here
$$
\gamma_{\sigma}^{(+-)}(q_{1},q_{2})=
\gamma_{\sigma}-
(\hat q_{1}-M)\frac{n^{-}_{\sigma}}{p^{-}_{D_{1}}+p^{-}_{D_{2}}}
+(\hat q_{2}-M)\frac{n^{+}_{\sigma}}{p^{+}_{D_{1}}+p^{+}_{D_{2}}}
$$
is the effective vertex  for the transition of two reggeized
quarks into
one gluon.

The induced vertex coming from the diagrams without particles in
$q_1^2$ or $q_2^2$ channels is
\begin{eqnarray} &&
\Delta_{\mu\nu}^{d_{1}d_{2}}(q_{1},q_{2})= (\hat
q_{2}-M)\frac{n^{+}_{\mu}n^{+}_{\nu}}
{p^{+}_{D_{1}}+p^{+}_{D_{2}}}\,
\left(\frac{a}{p_{D_{1}}^+}+\frac{b}{p_{D_{2}}^+}\right)
\, T^{d_{2}}T^{d_{1}}+ \nonumber \\
&&
+(\hat q_{1}-M)\frac{n^{-}_{\mu}n^{-}_{\nu}}
{p^{-}_{D_{1}}+p^{-}_{D_{2}}}\,
\left(\frac{c}{p_{D_{1}}^-}+\frac{d}{p_{D_{2}}^-}\right)
\, T^{d_{2}}T^{d_{1}} +
(p_{D_{1}}\leftrightarrow p_{D_{2}},\, d_{1}\leftrightarrow d_{2}).
\label{214}
\end{eqnarray}
The coefficients $a=0,b=1,c=1,d=0$
are fixed without any ambiguity  from the requirement of the
gauge invariance of the
amplitude.

Further, the effective amplitude for the  transition of the virtual
(reggeized)
quark (with momentum $q_1$) and gluon (with momentum $-q_2$) into
the real quark (with momentum $p_{D_1}$) and gluon (with momentum
$p_{D_2}$) is
\begin{eqnarray}
&& \frac{1}{2}g^{2}e^{*}_{D_2}\cdot \bar u_{D_1} \left(
\gamma^{(-)}_{\sigma}(p_{D_1}-q_{1},q_{1})\, (p_{D_1}-q_{1})^{-2}\!
\left(\gamma_{\mu\nu\sigma}(p_{D_2},q_{2})\, n^{+}_{\nu}-
q_{2}^{2}\frac{n^{+}_{\mu}n^{+}_{\sigma}}{p_{D_{2}}^+}\right)\!
[T^{c},T^{d_{2}}]- \right. \nonumber \\
&&\!-\gamma^{+}(\hat q_{1}-\hat p_{D_2}-M)^{-1}
\gamma^{(-)}_{\mu}(-p_{D_2},q_{1})\, T^{c}T^{d_2}\!-\!
\gamma_{\mu}(\hat q_{1}-\hat q_{2}-M)^{-1}
\gamma^{(-)}_{\sigma}(-q_{2},q_{1})\, n^{+}_{\sigma} T^{d_2}T^{c}\!-\!
\nonumber \\
&&\left. -\left[2(\hat q_{1}-M)
\frac{n^{-}_{\mu}}{p^{-}_{D_{2}}+q^{-}_{2}}
\left(\frac{T^{d_2}T^{c}}{q^{-}_{2}}+\frac{T^{c}T^{d_2}}{p^{-}_{D_{2}}}
\right) \right] \right)\, ,\!\!\!\!
\end{eqnarray}
where $\gamma^{+}\equiv \gamma_{\sigma}n^{+}_{\sigma}$ and
$c$ is the colour index of the virtual (reggeized) gluons.

At last the effective amplitude for the transition of the virtual
(reggeized)
quark (with  flavour
$F_{1}$) and anti-quark  (with
flavour
$F_{2}$) into the real quark $D_1$ and anti-quark $D_2$ is
\begin{eqnarray}
&&g^{2}\left(
(\bar u_{D_1}\gamma_{\sigma}T^{c}v_{D_2}) s_{2}^{-1}
\gamma_{\sigma}^{(+-)}(q_{1},q_{2})\,T^{c}
\cdot \delta_{F_{1}F_{2}}\delta_{D_{1}D_{2}}+
\right. \nonumber \\ && \left.
+(q_{1}-p_{D_1})^{-2}\,T^{c}
\gamma^{(+)}_{\sigma}(-p_{D_2}-q_{2},q_{2}) v_{D_2}
\otimes \bar u_{D_1}\gamma^{(-)}_{\sigma}(p_{D_1}-q_{1},q_{1})\,T^{c}
\cdot \delta_{F_{1}D_{1}}\delta_{F_{2}D_{2}}\right).
\end{eqnarray}

\section{Effective action for the reggeized
gluon and quark interactions}



The effective action
for the reggeized gluon interactions local in the
rapidity interval
$(y_0-\frac{\eta}{2},y_0+\frac{\eta}{2})$
($\eta\ll \ln\frac{s}{M^{2}}$) can be written as follows \cite{eff}
\begin{equation}
S_{G_{eff}}\!\!=\int\!\! d^{4}x \!\left( L_{QCD} +tr \,
\left((A_{+}(v_+)-A_{+})\partial^{2}_{\mu}A_{-}
+(A_{-}(v_-)-A_{-})\partial^{2}_{\mu}A_{+}
\right) \right) \,,
\label{32}
\end{equation}
where the usual QCD lagrangian is well known
\begin{eqnarray}
L_{QCD} =\frac{1}{2} \, tr\, G_{\mu \nu}^2+\bar\psi(i\hat D-M)\psi
\,,\,\,G_{\mu\nu}=\frac{1}{g}\left[D_{\mu},D_{\nu}\right] \,,\,\,
D_{\mu}=\partial _{\mu}+gv_{\mu}\,.
\end{eqnarray}
It is expressed in terms of the anti-hermitian gluon field $v_{\mu
}=-iv_{\mu}^{a}T^{a}$ ($T^a$ is the colour group generator in the
fundamental  representation) and the quark field $\psi$.
The composite field
$A_{\pm}(v_{\pm})$  contains an infinite number of terms
\begin{equation}
A_{\pm}(v_{\pm})=
\sum_{n=0}^{\infty}(-g)^{n}v_{\pm}
(\partial_{\pm}^{-1}v_{\pm})^{n}
\label{33}
\end{equation}
and can be written in terms of the covariant
derivative
\begin{eqnarray}
A_{\pm}(v_{\pm}) = -\frac{1}{g}\,\partial _{\pm}\,
\frac{1}{D _{\pm}}\, \partial _{\pm} \,1\,,
\end{eqnarray}
where $1$ is a unit matrix.

We can chose different definitions for the integral operators
$\partial_\pm ^{-1}$ to express $A_{\pm}(v_{\pm})$ through
$P$-ordered Wilson exponents with the integration contours displaced
along light cone lines.
The Feynman diagram prescription corresponds to the
symmetric form of $A_{\pm}(v_{\pm})$
\begin{eqnarray}
A_\pm (v_{\pm})=
-\frac{1}{g}\,\partial_\pm \,U(v_{\pm})\,,
\end{eqnarray}
\begin{eqnarray}
U(v_{\pm})=\frac{P}{2}
 \exp \left(
-\frac{g}{2}\int_{-\infty}^{x^{\pm}}\!\!
dx'^{\pm}v_{\pm}(x')
\right)+\frac{\bar{P}}{2}\exp \left(
\frac{g}{2}\int ^{\infty}_{x^{\pm}}\!\!
dx'^{\pm}v_{\pm}(x')
\right),\,\, U^+(v_\pm )=U^{T*}(v_\pm ).
\end{eqnarray}



The anti-hermitian fields $A_{\pm}=-iA_{\pm}^{a}T^{a}$ describe
the production and
annihilation of the reggeized gluons. Their
bare propagator is
\begin{equation}
\int\!\! d^{4}x e^{-ipx}
<A_{+}^{ay'}(x)A_{-}^{by}(0)>=
q_{\perp}^{-2}\cdot\theta(y'-y-\eta)\delta^{ab}\, .
\label{34}
\end{equation}
The reggeon fields  satisfy the kinematical constraints \cite{eff}
$$
\partial_{+}A_{-}=\partial_{-}A_{+}=0.
$$
Taking also into account the above prescription for $U(v_\pm)$ it
leads
to the followng expression for the reggeon-gluon interaction
term \cite{s2}
\begin{eqnarray}
S^{int}_G=-\frac{1}{g} \,tr\,\int d^2\,\vec{k}
\left( \int
_{-\infty}^{\infty}\,d\,x^+\,T(v_-)\,\partial^{2}_{\mu}A_{+}+\int
_{-\infty}^{\infty}
\,d\,x^- \,T(v_+)\,\partial^{2}_{\mu}A_{-}
\right)\,,
\end{eqnarray}
\begin{eqnarray}
 T(v_\pm )=\frac{1}{2} \left( P\, \exp \left(
-\frac{g}{2}\int_{-\infty}^{\infty}\!\!
dx'^{\pm}v_{\pm}(x')
\right)-\bar{P}\exp \left(
\frac{g}{2}\int ^{\infty}_{-\infty}\!\!
dx'^{\pm}v_{\pm}(x')
\right)\right)
\end{eqnarray}
and provides  the hermicity property of the effective
action.

Under the gauge transformations of the gluon and quark fields
\begin{eqnarray}
v_{\mu}\rightarrow \frac{1}{g} \,e^{\chi (x)}\,D_{\mu}\,e^{-\chi
(x)}\,,\,\,\psi \rightarrow e^{\chi (x)}\psi \,,\,\,
\bar{\psi} \rightarrow \bar{\psi}e^{-\chi (x)}
\end{eqnarray}
we have
\begin{eqnarray}
U(v_{\pm})\rightarrow e^{\chi}\,U(v_{\pm})\,,\,\,
U^+(v_{\pm})\rightarrow U^+(v_{\pm})\,e^{-\chi}\,,\,\,T(v_\pm
)\rightarrow T(v_\pm ).
\end{eqnarray}
Using the requirement, that for the gauge parameter falling at large
distances ($\chi (x) \rightarrow 0$, $x
\rightarrow \infty$) the
reggeon fields $A_{\pm}$ are invariant
\begin{eqnarray}
A_{\pm} \rightarrow A_{\pm}\,,
\end{eqnarray}
one can verify, that $S_{G_{eff}}$
is also gauge invariant \cite{eff}. Because $A_{\pm}$ belong to the
adjoint
representation of the gauge group, the action is invariant also under the
global gauge
transformations.

The action describing the gauge-invariant interaction of the gluon and
quarks
with  the reggeized quarks within the fixed interval of
rapidities $\eta$ can be written as follows
\begin{eqnarray}
 S_{Q_{eff}}\!\!=\int\!\! d^{4}x & \!\!\!\!
\left( {\bar a}_{-}(i\hat{\partial}-M)(a_{+}-U^{+}(v_+)\psi)+
{\bar a}_{+}(i\hat{\partial}-M)(a_{-}-U^{+}(v_-)\psi)+\right.
\nonumber \\
&\!\!\!\!\! \left. +({\bar a}_{+}-\bar\psi U(v_+))
(i\hat{\partial}-M)a_{-}+
({\bar a}_{-}-\bar\psi U(v_-))(i\hat{\partial}-M)a_{+})\right)\,,
\label{35}
\end{eqnarray}
where $a_+$, (${\bar a}_+$) are the fields describing the production
of (reggeized) quarks (anti-quarks) in the $t$-channel and  $a_+$, (${\bar
a}_+$) are the
corresponding fields
for their
annihilation (cf. \cite{s1}).

The reggeon fields satisfy the kinematical constraints
\begin{equation}
\partial_{+}a_{-}=\partial_{-}a_{+}=0,\quad
\partial_{+}\bar a_{-}=\partial_{-}\bar a_{+}=0,
\label{36}
\end{equation}

\begin{equation}
\gamma_{+}a_{-}=\gamma_{-}a_{+}=0,\quad
\bar a_{-}\gamma_{+}=\bar a_{+}\gamma_{-}=0\,.
\label{36a}
\end{equation}
The first constraint is similar to the analogous condition for
$A_\pm$ and means. that the light-cone components $q^+$ and
$q^-$ of the reggeon momentum $q$ are much smaller than the
corresponding components of the particles
emitting and absorpting the reggeon. The second constraint
is related with  analogous relations for the $\gamma$-matrix
light-cone components.

The reggeon fields are assumed to be gauge
invariant
\begin{eqnarray}
a_{\pm} \rightarrow a_{\pm}\,,\,\, \bar{a}_{\pm} \rightarrow \bar{a}_{\pm}
\end{eqnarray}
if the parameters $\chi (x)$
decrease at large $x$. They are transformed according to the
fundamental
representations under the global
colour rotations. It means, that $ S_{Q_{eff}}$ is invariant under local
and global gauge transformations.

The operators $1/{\partial_\pm}$ correspond to the
propagators of the virtual particles emitting the gluons within the given
interval of rapidities. Indeed, the factor $i\hat{\partial}-M$
standing in front of
the Wilson
exponents in the action cancels in the Feynman diagrams the
neighbouring propagator of the
reggeon
fields $a_\pm ,\,\bar{a}_\pm$, which gives a possibility to
interprete
$1/{\partial _\pm}$ as the colour particle propagators. Therefore the
analitic structure of the Feynman diagrams in the effective field theory
is the same as in the initial QCD. According to the
equations of motion for the guark field
\begin{eqnarray}
a_\pm  =  U^{+}(v_\pm )\,P_\pm \psi \,,\,\, P_\pm =\frac{1}{2}\,
\gamma _\mp \, \gamma _\pm
\end{eqnarray}
the reggeon field $a_\pm$ can be considered as a classical component
of
the corresponding quark field
coinciding with it in the perturbation theory $g\rightarrow 0$ or
in the light-cone gauge. Moreover, the independence of the physical
results from the intermediate rapidity parameter $\eta$ gives a
possibility to go in a continious way from the usual QCD
corresponding to $\eta =\ln s$
with the absence of the phase space for $A_\pm$ to the reggeon
field theory where $\eta$ is small (cf. \cite{s2}).

%

After the shift of the fermion fields $\psi\to\psi+a_{+}+a_{-}$,
$\bar\psi\to\bar\psi+\bar a_{+}+\bar a_{-}\,$
the bare propagators for the reggeized fermions can be written as
follows
\begin{eqnarray}
&& \int\!\! d^{4}x e^{-iqx}
<a_{+}^{y'}(x)\bar a_{-}^{y}(0)>=P_{+}
 (\hat q_{\perp}-M)^{-1}\cdot\theta(y'-y-\eta),
\nonumber \\
&& \int\!\! d^{4}x e^{-iqx}
<a_{-}^{y}(x)\bar a_{+}^{y'}(0)>=P_{-}
 (\hat q_{\perp}-M)^{-1}\cdot\theta(y'-y-\eta).
\label{37}
\end{eqnarray}

The total gauge-invariant action for reggeized gluon and quark
interactions is
\begin{eqnarray}
S_{{}_{eff}}=S_{G_{eff}}+S_{Q_{eff}}\,.
\end{eqnarray}
This action is local in the rapidity interval $(y_0-\eta,
\,y_0+\eta )$. The physical results do not depend on $\eta$ due
to the cancelation of this dependence in the two-dimensional
integrals of the reggeon field theory. The effective action
allows to reproduce easily the results of  calculations of all
known reggeon-particle amplitudes.
\section{Conclusion}
In this paper we constructed the Born amplitudes for the processes
of the  particle production in the quasi-multi-Regge
kinematics. To calculate two-loop corrections to the quark Regge
trajectory one can use the $t$- or $s$-channel unitarity conditions
(cf. \cite{b2}, \cite{b89}).
In both cases two- and three- particle intermediate states should be
taken into account. For two-particle contributions one should
find the elastic amplitudes in the one-loop apprximation, which
can be done with the use of the $t$-channel unitarity (see
\cite{b2}, \cite{b89}). For three-particle intermediate states in
the
$s$-channel there are three kinematical regions of integration
corresponding to the multi-Regge and quasi-multi-Regge kinematics.
In the first region one can use the results of LLA for the
production amplitudes and for two other regions the production
amplitudes are calculated in this paper. After the integration over
the phase space of the intermediate particle momenta and taking
into account two-particle contributions we should subtract the
leading term proportional to $\ln s$ and appearing from LLA for
the Regge trajectory. The contribution to the Regge trajectory from
the three-particle intermediate state in the $t$-channel does not
contain the divergency, corresponding to LLA. It appears only
in the two-particle contribution and should be subtructed to avoid
the double-counting.

To calculate the one-loop correction to the kernel of the integral
Bethe-Salpeter equation for the $t$-channel partial wave with
meson quantum numbers,
apart from the two-loop correction to the quark Regge
trajectory
and the reggeon-reggeon-particle-particle amplitude, calculated
in this paper, one should know also one-loop correction to the
reggeon-reggeon-particle, which can be found with the use of the
crossing channel unitarity relations and the results of this
paper (cf. \cite{b2}, \cite{b89}).

Another method of calculations of the next-to-leading corrections
to the quark Regge trajectory and to the integral kernel for the
Bethe-Salpeter equation is based on the above constructed  effective
action for interactions of the reggeized quark and gluons with the
usual QCD partons (cf. \cite{s2}). We
hope to return to these problems in our subsequent publications.

\[
\]

{\large{\bf Acknowledgements}} \newline

L.N. Lipatov thanks  the University Montpellier-2 for the support of
his stay in France where this work was finished. M.I. Vyazovsky is
grateful to the Petersburg Nuclear Physics Institute for awarding
him the International Gribov's prize for young theorists.


\newcommand{\Jour}[3]{{\bf #1} (#2), {#3}}


\newpage
\begin{thebibliography}{99}

     \bibitem{bfkl} V.S.Fadin, E.A.Kuraev, L.N.Lipatov
     $//$ Phys.Lett. \Jour{B60}{1975}{50} \\
                    E.A.Kuraev, L.N.Lipatov, V.S.Fadin
     $//$ ZhETF \Jour{71}{1976}{840}; \Jour{72}{1977}{377} \\
                    Ya.Ya.Balitskiy, L.N.Lipatov
     $//$ Yad. Fiz. \Jour{28}{1978}{1597}

     \bibitem{s2} L.N.Lipatov
     $//$ Physics Reports \Jour{286}{1997}{132}

     \bibitem{b2} V.S.Fadin, L.N.Lipatov $//$
     Nucl. Phys. \Jour{B406}{1993}{259}; \\
     V.S.Fadin, R.Fiore, M.I.Kotsky
     $//$ Phys.Lett. \Jour{B359}{1995}{181};
                     \Jour{B387}{1996}{593}

     \bibitem{b3} V.S.Fadin, L.N.Lipatov
     $//$ Phys.Lett. \Jour{B429}{1998}{127}; \\
                  G.Camici, M.Ciafaloni
     $//$ Phys.Lett. \Jour{B430}{1998}{349}

     \bibitem{b4} S.J.Brodsky, V.S.Fadin, V.T.Kim, L.N.Lipatov,
     G.B.Pivovarov,
     $//$ JETP Lett. \Jour{70}{1999}{155}\\
     V.T.Kim, L.N.Lipatov, G.B.Pivovarov
     $//$ hep-ph/9911228

     \bibitem{odd} L.N. Lipatov, $//$ Phys. Lett.
     \Jour{B309}{1993}{394};
     $//$ Sov. Phys. JETP Lett. \Jour{59}{1994}{571};\\
     L.D. Faddeev, G.P. Korchemsky, $//$ Phys. Lett.
     \Jour{B342}{1995}{311}\\
     J. Bartels, G.P. Vacca, L.N. Lipatov, $//$ Phys. Lett.
     \Jour{B477}{2000}{178}

     \bibitem{s1}  L.N.Lipatov
     $//$ Nucl.Phys. \Jour{B365}{1991}{614} \\
     R.Kirschner, L.N.Lipatov, L.Szymanovski
     $//$ Phys.Rev. \Jour{D51}{1995}{838}

     \bibitem{eff} L.N. Lipatov, $//$ Nucl.Phys.
\Jour{B425}{1995}{369}



     \bibitem{fsh} V.S.Fadin, V.E.Sherman
     $//$ ZhETF \Jour{72}{1977}{1640}

     \bibitem{ggmz}
    M.Gell-Mann, M.L.Goldberger, F.E.Low, E.Marx, P.Zachariasen $//$ \\
          Phys.Rev. \Jour{133B}{1964}{145}

     \bibitem{b89} L.N.Lipatov, V.S.Fadin
     $//$ Yad. Fiz. \Jour{50}{1989}{1141}


\end{thebibliography}
\end{document}